\documentclass[preprint,12pt,onecolumn]{emulateapj}
\usepackage{amstext}
\usepackage{apjfonts}
\usepackage{amsmath}

\begin{document}
\pdfoutput=1

\title{$\lowercase{e}^\pm$ pair loading and the origin of the upstream
magnetic field in GRB shocks}

\author{Enrico Ramirez-Ruiz\altaffilmark{1,2}, Ken-Ichi
  Nishikawa\altaffilmark{3} and Christian B. Hededal\altaffilmark{4}}
\altaffiltext{1}{Institute for Advanced Study, Einstein Drive,
  Princeton, NJ 08540, USA} \altaffiltext{2}{Department of Astronomy
  and Astrophysics, University of California, Santa Cruz, CA 95064,
  USA}\altaffiltext{3}{National Space Science and Technology Center,
  Huntsville, AL 35805, USA}\altaffiltext{4}{Dark Cosmology Center,
  Niels Bohr Institute, Juliane Maries Vej 30, 2100 Copenhagen,
  Denmark}

\begin{abstract} 
We investigate here the effects of plasma instabilities driven by
rapid $e^{\pm}$ pair cascades, which arise in the environment of GRB
sources as a result of back-scattering of a seed fraction of the
original spectrum. The injection of $e^\pm$ pairs induces strong
streaming motions in the ambient medium. One therefore expects the
pair-enriched medium ahead of the forward shock to be strongly sheared
on length scales comparable to the radiation front thickness. Using
three-dimensional particle-in-cell simulations, we show that plasma
instabilities driven by these streaming $e^\pm$ pairs are responsible
for the excitation of near-equipartition, turbulent magnetic
fields. Our results reveal the importance of the electromagnetic
filamentation instability in ensuring an effective coupling between
$e^\pm$ pairs and ions, and may help explain the origin of large
upstream fields in GRB shocks.
\end{abstract}
 
\keywords{gamma rays: bursts -- instabilities -- magnetic fields --
    plasmas -- shock waves}

\section{Introduction}
More than three decades ago, it was pointed out that $\gamma$-rays
produced in sufficiently luminous and compact astrophysical sources
would create $e^{\pm}$ pairs by collisions with lower energy photons:
$\gamma\gamma \rightarrow e^{+}e^{-}$ \citep{jel66}. This mechanism
both depletes the escaping radiation and also changes the composition
and properties of the radiating gas through the injection of new
particles. An approximate condition for such pair creation to become
significant is that a sizable fraction of the radiation from the
object be emitted above the electron mass energy, $m_e c^2=$ 511 keV,
and that the compactness of the source,
\begin{equation}
l={L\sigma_T \over 4\pi r_l m_e c^3},
\end{equation}
exceeds 1, where $L$ is the total luminosity, and $r_l$ is the
characteristic source dimension. When $l \sim 1$, a photon of energy
$\epsilon=h\nu/m_{\rm e}c^2 \sim 1$ has an optical depth of unity for
creating an $e^{\pm}$ pair.

As an illustration consider a spherical source with a spectrum
emitting a power $L$ in each decade of frequencies, $l$ would be $>1$
for $\gamma$-rays of energy $\epsilon$ if the radius of the source
satisfies
\begin{equation}
r_l \leq 10^{17} \epsilon \left({L \over 10^{49} {\rm erg\;s}^{-1}}\right){\rm
cm}.
\label{eqn:ssize}
\end{equation}
The relevant values of $L$ range from $10^{50}-10^{52}$ erg s$^{-1}$
for GRBs. In addition, the short spikes observed in the high energy
light curves suggest that GRBs dissipate a significant fraction of
their energy at $ r \ll r_l$, so that the source is indeed so small
that equation (\ref{eqn:ssize}) is easily satisfied. This argument is,
however, only applicable to $\gamma$-ray photons emitted isotropically
and is alleviated when the radiating source itself expand at a
relativistic speed \citep{p99}. In this case, the photons are beamed
into a narrow angle $\theta \sim 1/\Gamma$ along the direction of
motion, and, as a result, the threshold energy for pair production
within the beamed is increased to $\epsilon \sim \Gamma$.

The non-thermal spectrum of GRB sources is therefore thought to arise
in shocks which develop beyond the radius at which the relativistic
fireball has become optically thin to $\gamma\gamma$ collisions
\citep{p99}. However, the observed spectra are hard, with a
significant fraction of the energy above the $\gamma\gamma \rightarrow
e^{+}e^{-}$ formation energy threshold, and a high compactness
parameter can result in new pairs being formed outside the originally
optically thin shocks responsible for the primary radiation
\citep{tm00}. Radiation scattered by the external medium, as the
collimated $\gamma$-ray front propagates through the ambient medium,
would be decollimated, and, as long as $l\gtrsim 1$, absorbed by the
primary beam. An $e^\pm$ pair cascade can then be produced as photons
are back-scattered by the newly formed $e^\pm$ pairs and interact with
other incoming seed photons \citep{tm00,b02,b05,m01,rr02,li03,kp04}.

In this paper, we consider the plasma instabilities generated by rapid
$e^\pm$ pair creation in GRBs. The injection of $e^\pm$ pairs induces
strong streaming motions in the ambient medium ahead of the forward
shock. This sheared flow will be Weibel-like unstable \citep{ml99},
and if there is time before the shock hits, the resulting plasma
instabilities will generate sub equipartition quasi-static long-lived
magnetic fields on the collisionless temporal and spatial scales
across the $e^\pm$ pair-enriched region. This is studied in \S
\ref{sim} using three-dimensional kinetic simulations of monoenergetic
and broadband pair plasma shells interpenetrating an unmagnetized
medium.  The importance of the electromagnetic filamentation
instability in providing an effective coupling between $e^\pm$ pairs
and ions is investigated in \S \ref{load}, while a brief description
of the numerical methods and the initial models is giving in \S
\ref{nm}.  The implications for the origin of the upstream magnetic
field, in particular in the context of constraints imposed by
observations of GRB afterglows, are discussed in \S \ref{con}.

\section{Pair Loading and  Magnetic Field Generation}\label{load}
Given a certain external baryon density $n_p$ at a radius $r$ outside
the shocks producing the GRB primordial spectrum, the initial Thomson
scattering optical depth is $\tau \sim n_p\sigma_T r$ and a fraction
$\tau$ of the primordial photons will be scattered back, initiating a
pair $e^\pm$ cascade. Since the photon flux drops as $r^{-2}$, for a
uniform (or decreasing) external ion density most of the scattering
occurs between $r$ and $r/2$, and the scattering and pair formation
may be approximated as a local phenomenon.

\begin{figure*}
\center{\includegraphics[scale=1.8]{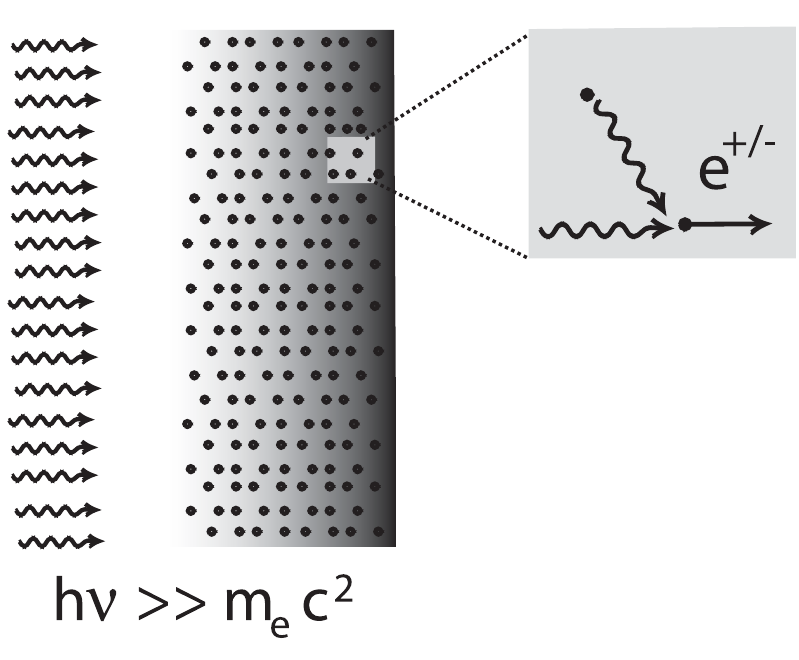}}
\caption{Schematic plot of $e^\pm$ pair cascades triggered by the
  back-scattering of seed $\gamma$-ray photons on the external
  medium.}
\label{fig1}
\end{figure*}

Consider an initial input GRB radiation spectrum of the form
$F(\epsilon)=F_\beta (\epsilon/\epsilon_\beta)^{-\alpha}$ for
$\epsilon >\epsilon_\beta$, where $\epsilon_\beta \sim 0.2-1.0$ is the
break energy above which the spectral index $\alpha\sim 1-2$ (for the
present purposes the exact low energy slope is unimportant). Radiation
scattered by the external medium is therefore decollimated, and then
absorbed by the primary beam.The impact of such process strongly
depends on how many photons each electron is able to scatter
\citep{b02}
\begin{equation}
\lambda_T\sim 4 \times 10^{8} \left({r_l \over 10^{15}\;{\rm
cm}}\right)^2 \left({L \over 10^{51}\;{\rm erg\;s^{-1}}}\right)\;{\rm cm}
<\Delta,
\end{equation}
where $\lambda_T=1/(n_\gamma \sigma_T) \approx c/(F_\beta\sigma_T)$ is
the electron mean free path and $n_\gamma$ is the photon density.  The
photons forced out from the beam by the electrons can then produce
$e^\pm$ pairs as they interact with incoming photons
(Fig.~\ref{fig1}), so that a large number of scatterings implies a
large number of pairs created per ambient electron.

At some distance $\varpi\leq \Delta$ from the leading edge of the
radiation front, the number of photons scattered by one ambient
electron is $\sim \varpi/ \lambda_T$ and a fraction $\sim \varpi/
\lambda_{\gamma\gamma}$ of these photons are absorbed \cite{b02}. One
$e^\pm$ pair per ambient electron is injected when $\varpi =
(\lambda_T\lambda_{\gamma\gamma})^{1/2} \approx \lambda_T
\epsilon_{\rm th}^{\alpha/2}/[\Pi(\alpha)]^{1/2} \geq \lambda_T$,
where $\Pi(\alpha)=2^{-\alpha}(7/12)(1+\alpha)^{-5/3}$ \citep{sve87}
and $\epsilon_{\rm th}$ is the photon threshold energy for $e^\pm$
formation.  For $1.5 \lesssim \alpha \lesssim 2$, one has $15\lambda_T
\lesssim \varpi_l \lesssim 25\lambda_T$, so that pair creation
substantially lags behind electron scattering \citep{b02}.

Most of the momentum deposited through this process involves the
side-scattering of very soft photons, which collide with hard
$\gamma$-rays to produce energetic (and almost radially moving) pairs.
A photon of energy $\epsilon_r \ll 1$ that is side-scattered through
an angle $\sim \theta_r$ creates a pair if it collides with another
photon with energy exceeding $\epsilon_{\rm th} \sim 4(\theta^2_r
\epsilon_r)^{-1}$. The injected pair will be relativistic with
$\gamma_\pm \sim \epsilon_{\rm th} \geq 1$. The distribution of
injected pairs is therefore directly determined by the high energy
spectral index $\alpha$. This motivates our study in \S \ref{sim} of
radially streaming, relativistic pair plasma shells with a broadband
kinetic energy distribution interpenetrating an unmagnetized medium.

This pair-dominated plasma, as long as its density $n_\pm \lesssim
n_p(m_p/2m_e)$, is initially held back by the inertia of its
constituent ions, provided that the pairs remain coupled to the
baryons \citep{tm00,b02}. The latter is likely to be the case in the
presence of weak magnetic fields \citep{tm00}. In the absence of
coupling, the pair density would not exponentiate, mainly due to the
$(1-\beta)$ term in the scattering cross section, where $\beta=v/c$.
Instabilities caused by pair streaming relative to the medium at rest,
as we argue in \S \ref{sim}, are able to generate long-lived magnetic
fields on the collisionless temporal and spatial scales, which are
modest multiples of the electron plasma frequency, $\omega_e = (4\pi
e^2 n_e/ m_e)^{1/2}$, and the collisionless skin depth,
\begin{equation}
\lambda_e = {c\over \omega_e}\sim 5 \times 10^{5}\left({n_e \over
1\;{\rm cm^{-3}}}\right)^{-1/2} \ll \lambda_T <
\lambda_{\gamma\gamma}.
\end{equation}
Here $n_e$ is the electron number density. Plasma instabilities
therefore evolve faster than the time between successive scatterings
and much before the scattered photons are absorbed by the primary
radiation \citep{b02}. In the presence of transverse magnetic field
$B$ the pairs gyrate around field lines on the Larmor time,
$\omega_B^{-1}= m_e c /(B e)$. The net momentum of the $e^\pm$ pairs
is thus efficiently communicated to the medium
\citep{f04,hn05}. Magnetic coupling may dominate if $\omega_B >
\omega_e $, which requires $B^2/4\pi > n_e m_e c^2$ \citep{b02}. In
the sections that follows we present a quantitative discussion of the
effects of plasma instabilities driven by rapid $e^{\pm}$ pair
injection.

\section{Simulation Model}\label{nm}
Here we illustrate the main features of the collision of an $e^\pm$
pair plasma shell into an unmagnetized medium, initially at rest,
using a modified version of the PIC code TRISTAN -- first developed by
\cite{b93} and most recently updated by \cite{n05,n06}. The
simulations were performed on a 85 $\times$ 85 $\times$ 640 grid (the
axes are labeled as x, y, z) with a total of $3.8 \times 10^8$
particles with periodic ([x,y] plane) and radiative (z direction)
boundary conditions. In physical units, the box size is 8.9 $\times$
8.9 $\times$ 66.7 $(c/\omega_e)^3$, and the simulations ran for $60
\omega_e^{-1}$.

In the simulations, a quasi-neutral plasma shell, consisting of
$e^\pm$ pairs and moving with a bulk momentum $u_z = \gamma_0 v_z/c$
along the $z$ direction, penetrates an ambient plasma initially at
rest. Here $\gamma_0$ is the initial Lorentz factor of the pairs and
$v_z$ is the bulk velocity of the shell along $z$. $e^\pm$ pairs are
continuously injected at $z = 2.6\lambda_e = 25 \Lambda$, where
$\Lambda=\lambda_e/9.6$ is the grid size. The ion-to-electron mass
ratio of the ambient plasma is set to $m_i/m_e=20$, with both plasma
populations having a thermal spread with an electron rms velocity
$v_{\rm th}/c = 0.1$. For completeness, a purely $e^\pm$ pair ambient
plasma (i.e., a mass ratio $m_i/m_e$ of 1) is also studied, with a
thermal electron velocity $v_{\rm th}/c = 0.1$. The shell and the
ambient plasma have a density ratio of $0.75$. Two different bulk
Lorentz factor configurations for the injected (cold) $e^\pm$ pairs
are considered: a monoenergetic ($u_z = 12.5$, $v_{\rm th}/c = 0.01$)
and a broadband distribution ($3.0 \le u_z\le 30.0, v_{\rm th}/c =
0.01$). Both distributions have similar kinetic energy contents and
plasma temperatures (Fig.~\ref{fig2}).

\section{Results and Interpretation}\label{sim}
Much of our effort in this section is dedicated to investigate the
ability of $e^\pm$ pair loading to alter the physical parameters
characterizing the ambient medium. Some of the questions at the
forefront of attention include the level of the (electro) magnetic
field generated via plasma instabilities, as well as the saturated
state of the particles and fields. We address all of these issues
here. We consider both $e^\pm$ pair and electron-ion plasmas as
possible ambient media. In the dimensional units used, the results for
$e^\pm$ pair ambient plasmas are equally applicable to
interpenetrating proton/antiproton plasma shells.

\subsection{$e^\pm$ Pair Ambient Plasma}\label{pairs}
As pointed out in \cite{s03}, encountering the medium at rest, the
incoming pairs are rapidly deflected by field fluctuations growing
because of the two-stream instability \citep{ml99,p01,g01}. A large
number of oppositely directed current filaments are generated, as
illustrated in Fig.~\ref{fig3}, which in turn will generate
inhomogeneities in the magnetic field (predominantly in the x-y
plane). This configuration is unstable because opposite currents repel
each other, whereas like currents are attracted to each other and tend
to coalesce and form larger current filaments. During the linear stage
of the instability, the rapid generation of a large number of randomly
distributed current and magnetic filaments is observed (with a
correlation length $\lambda_J \approx 2^{1/4} \gamma_{\rm th}
(c/\omega_e)=1.2 \lambda_e=10\Lambda$) , which in turn is accompanied
by a expeditious production of a strong magnetic field
(Fig.~\ref{fig3}). The magnetic field energy density reaches 0.9\% of
the initial total kinetic energy ($\epsilon_p$) in the broadband case
and 1.2\% in the monoenergetic one (i.e., $\epsilon_B=\int (B^2 d
V/8\pi)/\epsilon_p$=0.009\;{\rm and}\; 0.012, respectively).

As the instability enters the saturation phase, these initially
randomly oriented filaments begin to interact with each other and
merge in a race where larger electron channels consume smaller
neighboring channels. In this manner, the transverse magnetic field
grows in strength. The instability then saturates and the energy in
the magnetic field decays. The strong decrease in the magnetic field
energy density is associated with a topological adjustment in the
structure of currents and fields \citep{s03}, and is mostly linked to
a reduction of the field's volume filling factor.  These distinctive
features are present in both the monoenergetic and broadband cases
(Fig.~\ref{fig3}). However, there are some clear contrasts.  For a
plasma shell with a broadband distribution of initial Lorentz factors,
transverse energy spreading happens over a variety of timescales. This
can be understood as follows. The instability arises from the free
streaming of particles, with a corresponding linear growth rate
scaling with $\gamma_0^{-1/2}$ \citep{ml99}.  Our experiments show
that this does indeed happen; the continuous injection of a broadband
distribution of momentum leads to a diversity of linear growth
rates. As the field amplitude grows, the transverse deflection of
particles increases (with decreasing $\gamma_0$), and thus free
streaming across the field is suppressed preferentially for the mildly
relativistic pairs. The magnetic field energy grows in the early
stages by slowing down the pair plasma shell.

\begin{figure*}
\center{\includegraphics[scale=0.9]{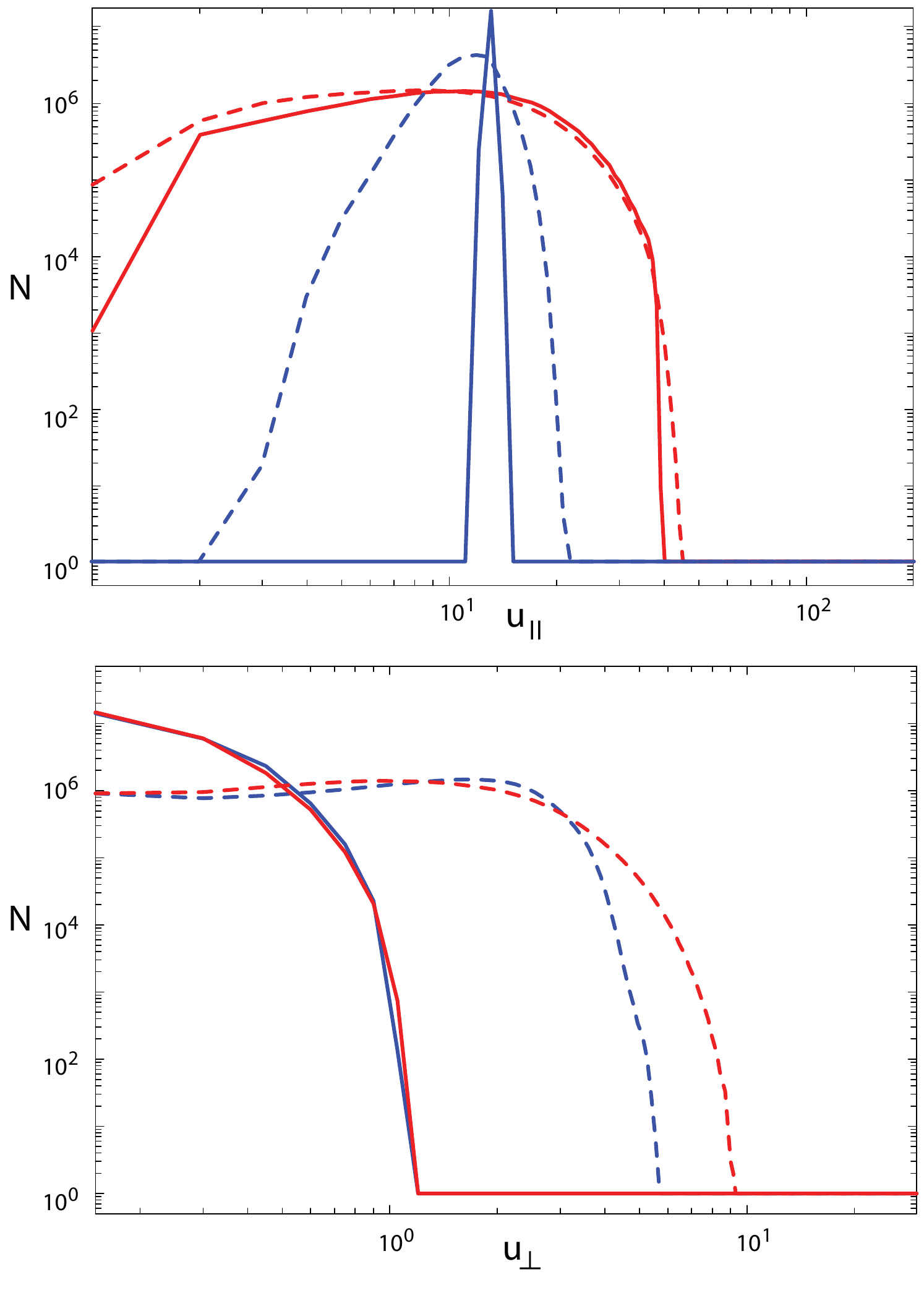}}
\caption{$e^\pm$ pair momentum distribution functions at $t=20.8
  \omega_e^{-1}$ ({\it solid} curves) and $t=59.8\omega_e^{-1}$ ({\it
    dotted} curves).  The {\it blue} ({\it red}) curves are for a
  monoenergetic (broadband) initial momentum distribution
  function. The {\it top} ({\it bottom}) panel show the distribution
  functions for particles moving parallel (perpendicular) to the
  injected $e^\pm$ pair plasma. }
\label{fig2}
\end{figure*}

\begin{figure*}
\center{\includegraphics[scale=1.0]{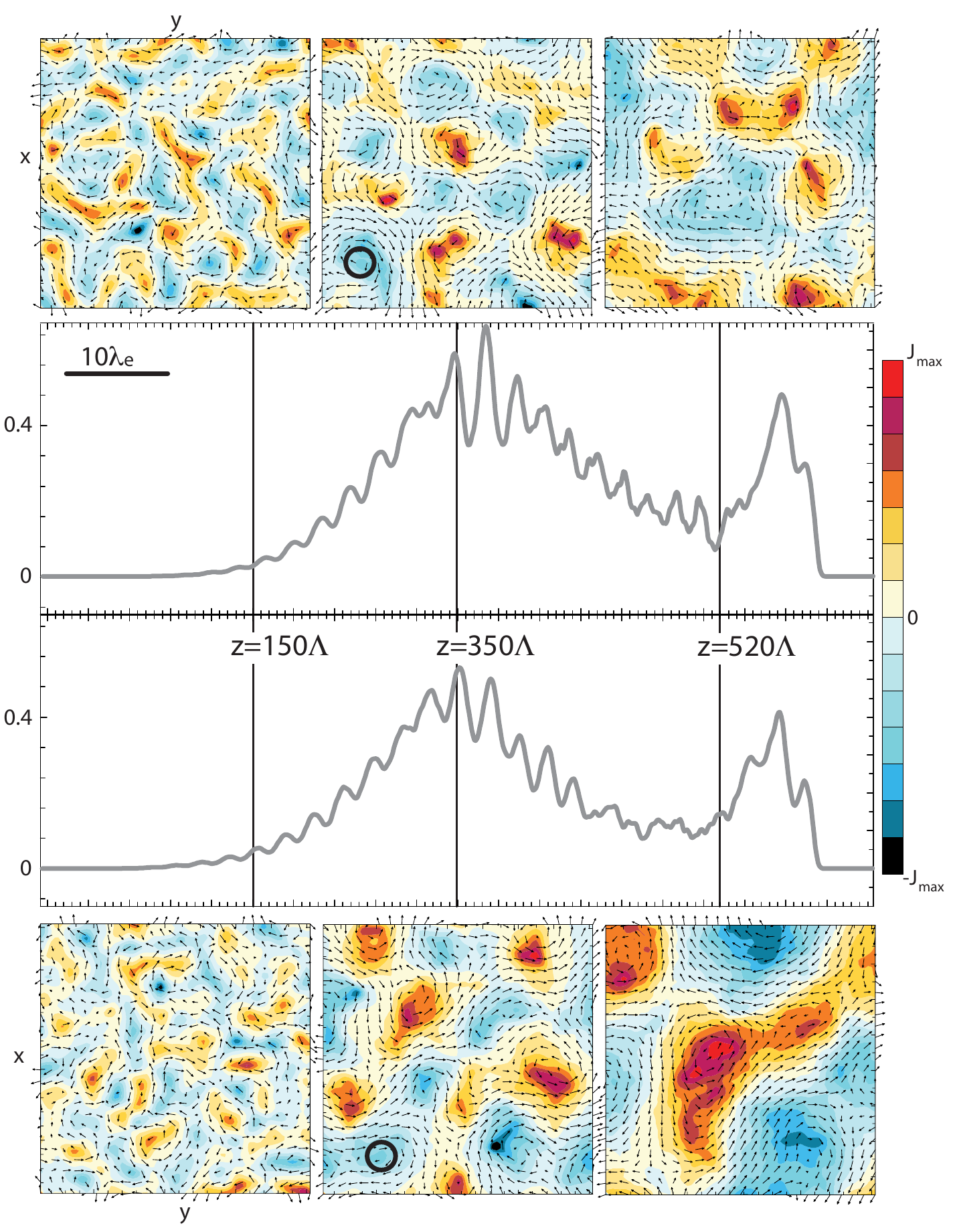}}
\caption{Growth of the two-stream instability at time
  $t=59.8\omega_e^{-1}$ for the two different momentum distribution
  function shown in Fig.~\ref{fig2}. Here we show the average
  transverse magnetic field amplitude (in simulation units) in the
  {\it x-y} plane as a function of $z$ ({\it solid} curves). The
  various panels show the {\it (x,y)} components of the magnetic field
  and the $z$ component of the current density ($J_z$) in different
  subsections of the computational box. The {\it top} ({\it bottom})
  panels are for a simulation in which the initial momentum
  distribution function magnetic field is monoenergetic
  (broadband). Color bar gives the amplitude of $J_z=[J_{\rm
      max},-J_{\rm max}]$ in simulation units for : $J_{\rm
    max}=20.04$ (25.57) for $z$=150, $J_{\rm max}=99.5$ (67.12) for
  $z$=350, and $J_{\rm max}=35.41$ (29.93) for $z$=520. The expected
  correlation length, $\lambda_J$, of the randomly distributed current
  and magnetic filaments is plotted in the $z$=350 subsections.}
\label{fig3}
\end{figure*}

\begin{figure*}
\center{\includegraphics[scale=1.0]{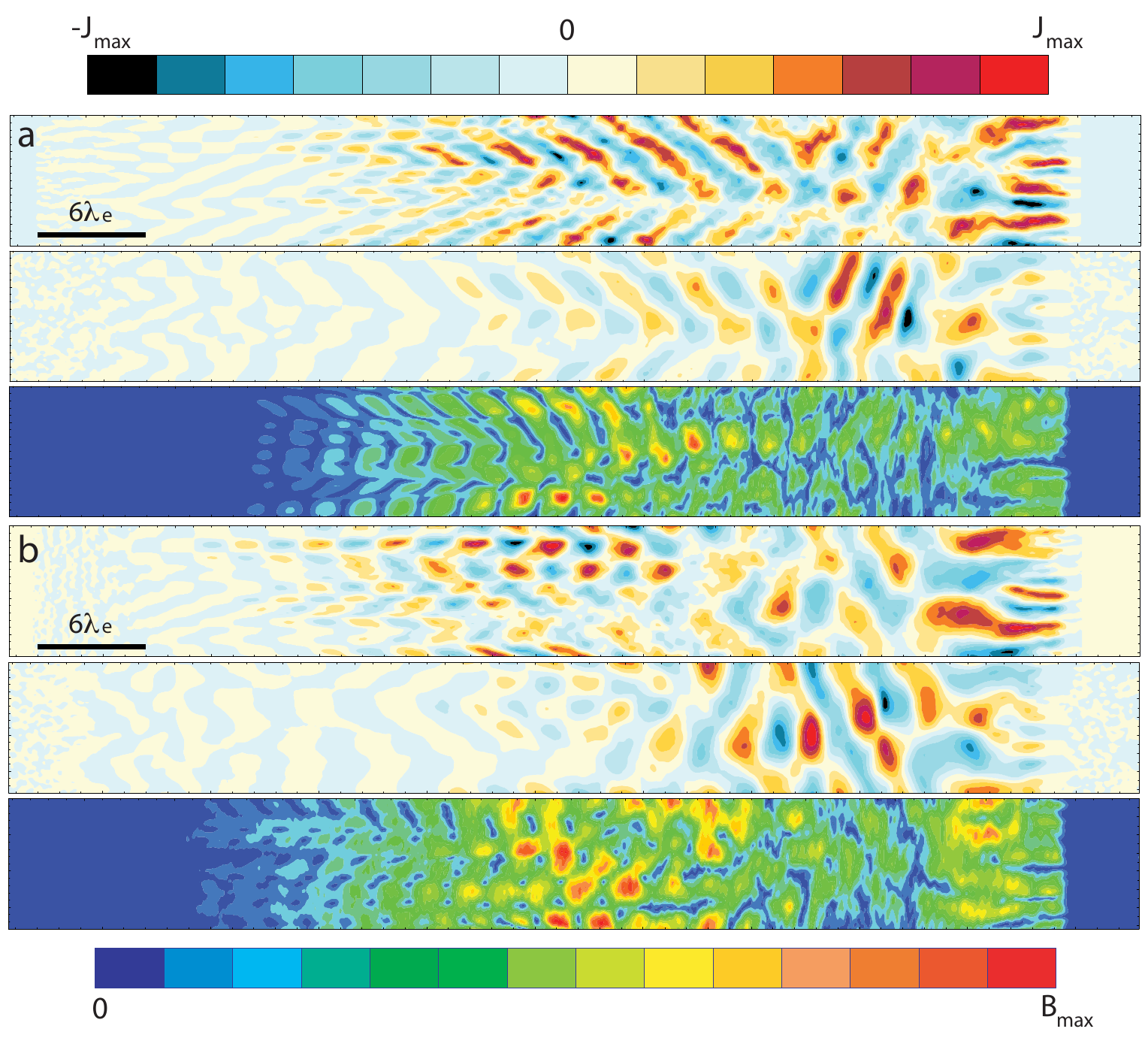}}
\caption{Current filamentation and magnetic field energy density after
  saturation: $t=59.8\omega_e^{-1}$. The amplitude of the average (in
  the {\it x-y} plane) transverse magnetic field, $B_\perp$ is plotted
  (in simulation units) as a function of $z$ together with the average
  $z$ component of the current density, $J_z$, for both injected ({\it
    top}) and ambient medium ({\it bottom}) $e^\pm$ pairs . The {\it
    a} ({\it b}) panels are for a simulation in which the initial
  momentum distribution function is monoenergetic (broadband). Color
  bar gives the amplitude of $J_z=[J_{\rm max},-J_{\rm max}]$ and
  $B_\perp=[0,B_{\rm max}]$ in simulation units: $J_{\rm max}=18.5$
  (17.3) for the injected $e^\pm$ pairs, $J_{\rm max}=15.9$ (13.9) for
  the ambient $e^\pm$ pairs and $B_{\rm max}=1.8$ (1.2).}
\label{fig4}
\end{figure*}

Saturation is then achieved by the combination of transverse energy
spreading and the generation of near-equipartition magnetic fields
\citep{s03}. In the broadband case, transverse energy spreading
happens more swiftly for mildly relativistic pairs, and, as a
consequence, $B_{\rm sat}$ is decreased. This follows directly from
\citep{ml99} the scaling of $B_{\rm sat}$ with $v_{\rm th}^{-1}$ (or
in this case, $v_{\perp}^{-1}$). As can be clearly seen in
Fig.~\ref{fig4}, the ambient $e^\pm$ pairs are more effectively heated
in the broadband case. After saturation, which takes place earlier for
mildly relativistic pairs, the energy stored in the magnetic field is
transferred back to the plasma particles, leading to strong heating
and the generation of a high-energy tail in the distribution
(Fig.~\ref{fig2}). The $e^\pm$ pairs are nonetheless expected to
thermalize, given sufficient space and time \citep{s06}.

It is not surprising that the linear growth rate saturates at
comparable timescales (Fig.~\ref{fig2}), since in the broadband
distribution, the energy averaged Lorentz factor, $\bar{\gamma_0}$, is
not that different from $\gamma_0$ in the monoenergetic
case. Fig.~\ref{fig3} also shows a transient, electromagnetic
precursor produced as the first injected $e^\pm$ pairs stream through
the ambient medium unhampered. The saturation level of the magnetic
field in the precursor region is expected to occur over longer
distances. This is because the linear growth rate of the
electromagnetic filamentation instability in the precursor region is
greatly decreased as a result of $e^\pm$ pairs being continuously
depleted from the shell's front \citep{s06}.

\subsection{Electron-Ion Ambient Plasma}\label{ion}
\begin{figure*}
\center{\includegraphics[scale=1.3]{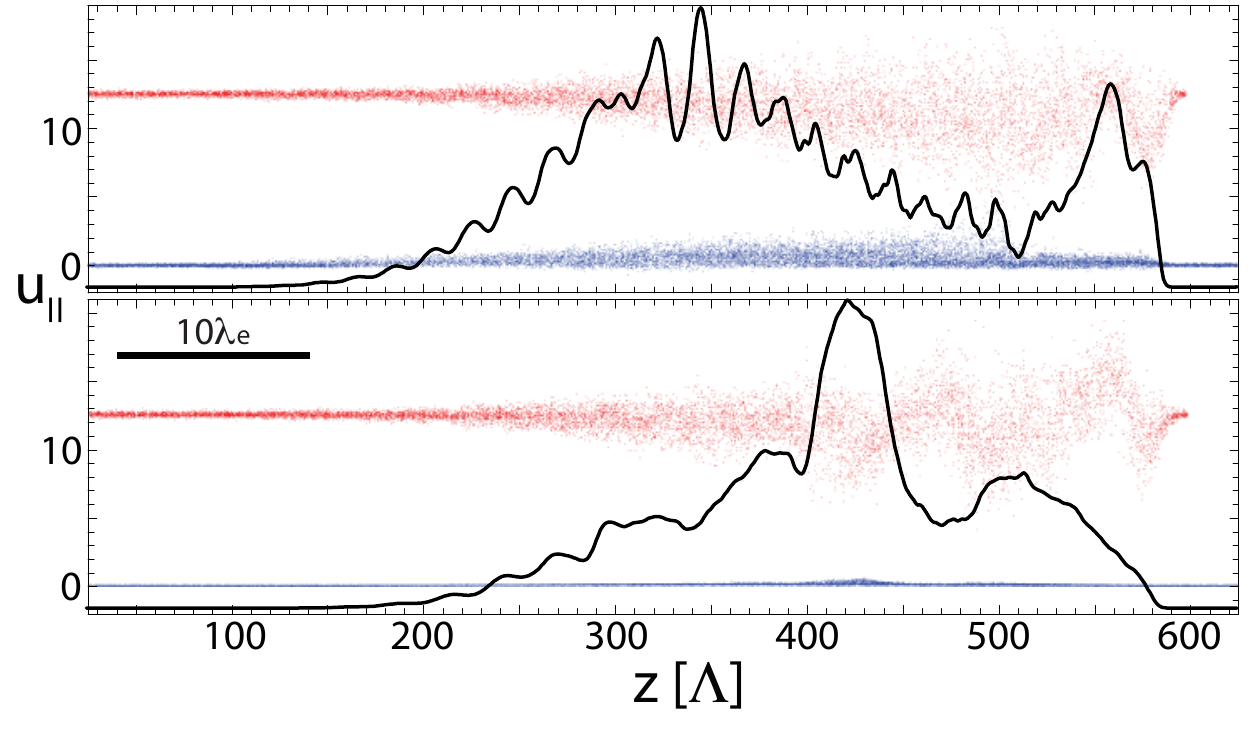}}
\caption{Longitudinal heating and acceleration, illustrated by changes
  in $u_\parallel$ for both injected (monoenergetic) $e^\pm$ pairs
  (red) and ambient plasma (blue). The {\it bottom} ({\it top}) panels
  are for a simulation in which the ambient medium is composed of
  electrons and ions ($e^\pm$ pairs). In the {\it top} panel, the
  ambient positrons are initially at rest but are strongly accelerated
  by the jet. In the {\it bottom} panel, the ions, being heavier than
  the pairs, remain clearly separated in phase space and are only
  slowly heated.  Also shown are the average transverse magnetic field
  amplitude (in arbitrary units) in the {\it x-y} plane as a function
  of $z$ ({\it solid} curves).}
\label{fig5}
\end{figure*}

Encountering the medium at rest, the incoming $e^\pm$ pairs are
rapidly deflected by field fluctuations growing because of the
two-stream instability. The initial perturbations grow nonlinear as
the deflected $e^\pm$ pairs collect into first caustic surfaces and
then current channels. The resultant cylindrical magnetic fields cause
mutual attraction between currents forcing like currents to approach
each other and merge. As a result, the magnetic field grows in
strength. This continues until the fields grow strong enough to
deflect the much heavier ions \citep{f04}. Fig.~\ref{fig5} shows that
this happens over $\sim 20$ electron skin depths from around $z=300$.
As illustrated in Fig.~\ref{fig5}, the ions stay clearly separated in
phase space and are only slowly heated. In the presence of ions, the
incoming $e^\pm$ pairs will drive higher levels of saturated B-field
(Fig.~\ref{fig5}), by a factor of $(m_i/m_e)^{1/2}$, albeit on a
longer timescale: the magnetic field energy density in the
monoenergetic case reaches 6.1\% of the initial total kinetic energy
in the electron-ion ambient plasma case and 1.2\% in the $e^\pm$ pair
one. This is due to the massive ion bulk momentum constituting a vast
energy reservoir for particle heating.

\begin{figure*}
\center{\includegraphics[scale=0.65]{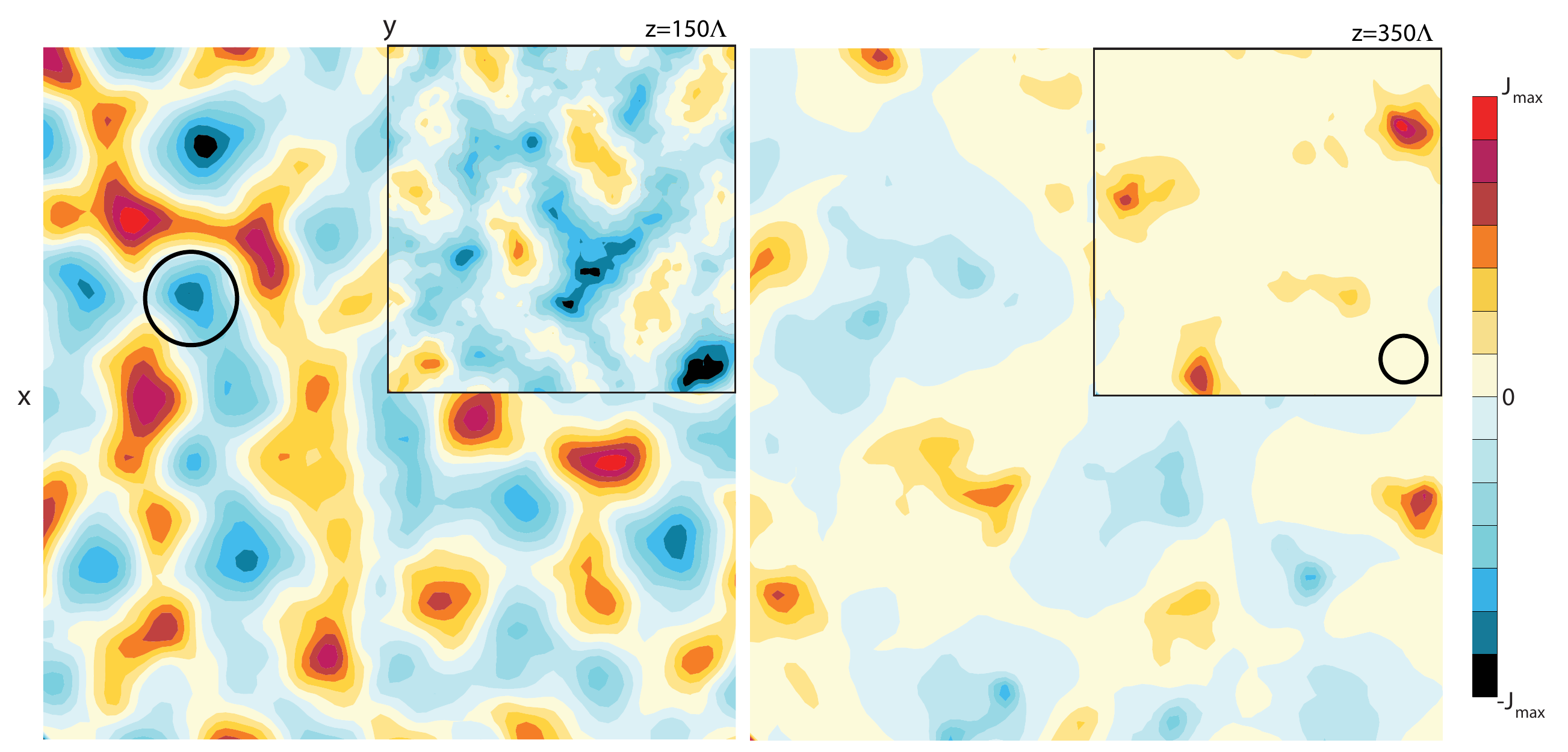}}
\caption{Longitudinal current densities in the counterstreaming
  plasmas. The various panels show the $z$ component of the current
  density, $J_z$, in different subsections of the computational box
  for $e^\pm$ pairs injected with a monoenergetic distribution. The
  small insets show the ion current in the same plane. Color bar gives
  the amplitude of $J_z=[J_{\rm max},-J_{\rm max}]$ in simulation
  units for $e^\pm$ pairs (ions) : $J_{\rm max}=20.04$ (0.1) for
  $z$=150 and $J_{\rm max}=173.3$ (8.1) for $z$=350. The expected
  correlation length, $\lambda_J$, of the randomly distributed current
  and magnetic filaments is plotted in the $z$=150 and $z$=350
  subsections.}
\label{fig6}
\end{figure*}

In the case of an electron-ion ambient plasma, the ion channels are
subjected to a similar growth mechanism as the positrons, but at a
slower rate. When ion channels grow sufficiently powerful, they begin
to experience Debye shielding by the electrons, which by then have
been significantly heated by scattering on the increasingly larger
electromagnetic field structures.  The temporal development of the
$e^\pm$ pair and ion channels is illustrated in Fig.~\ref{fig6}. The
large random velocities of the electron population allow the
concentrated ion channels to keep sustaining strong magnetic fields.

\begin{figure*}
\center{\includegraphics[scale=0.9]{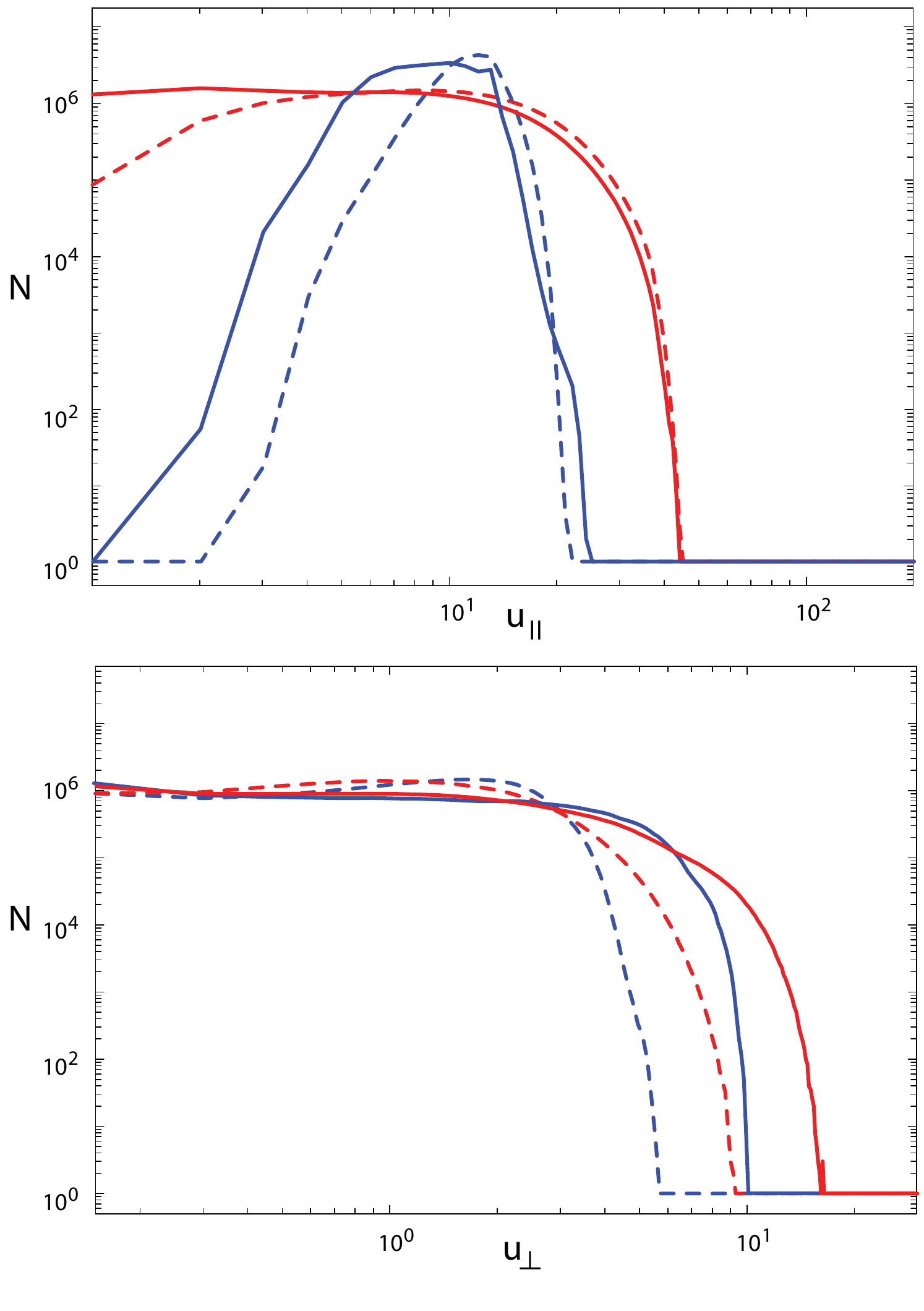}}
\caption{$e^\pm$ pair momentum distribution functions at
  $t=59.8\omega_e^{-1}$.  The {\it blue} ({\it red}) curves are for a
  monoenergetic (broadband) initial momentum distribution
  function. The {\it solid} ({\it dashed}) curves are for $e^\pm$
  pairs injected into a electron-ion ($e^\pm$ pair) plasma medium.}
\label{fig7}
\end{figure*}

\begin{figure*}
\center{\includegraphics[scale=1.2]{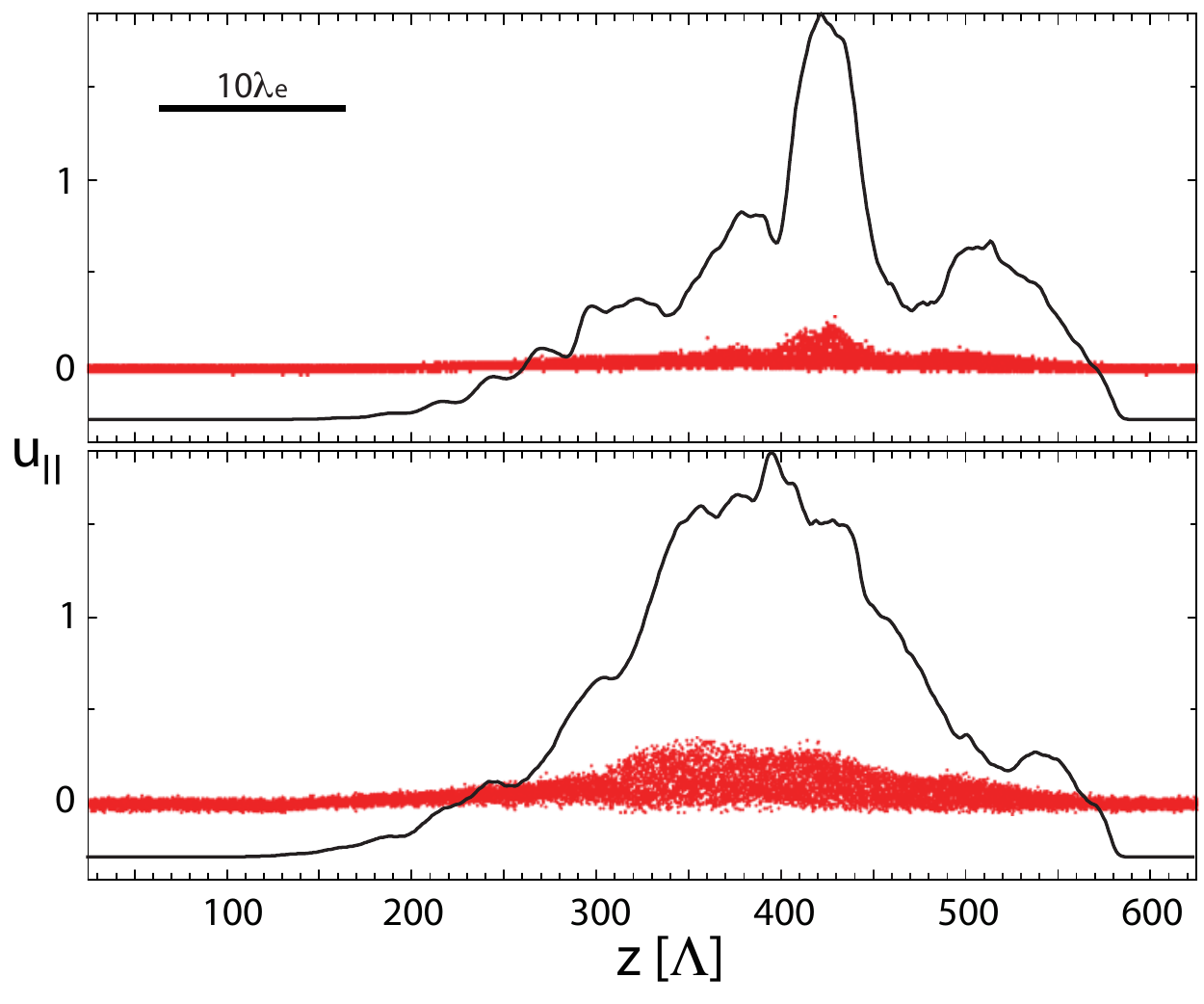}}
\caption{Longitudinal heating and acceleration of ambient ions. The
  {\it top} ({\it bottom}) panels are for a simulation in which the
  initial momentum distribution function of the injected $e^\pm$ pairs
  is monoenergetic (broadband) and the ambient medium is composed of
  electrons and ions.  Also shown is the average transverse magnetic
  field amplitude (in arbitrary units) in the {\it x-y} plane as a
  function of $z$ ({\it solid} curves).}
\label{fig8}
\end{figure*}

\begin{figure*}
\center{\includegraphics[scale=1.0]{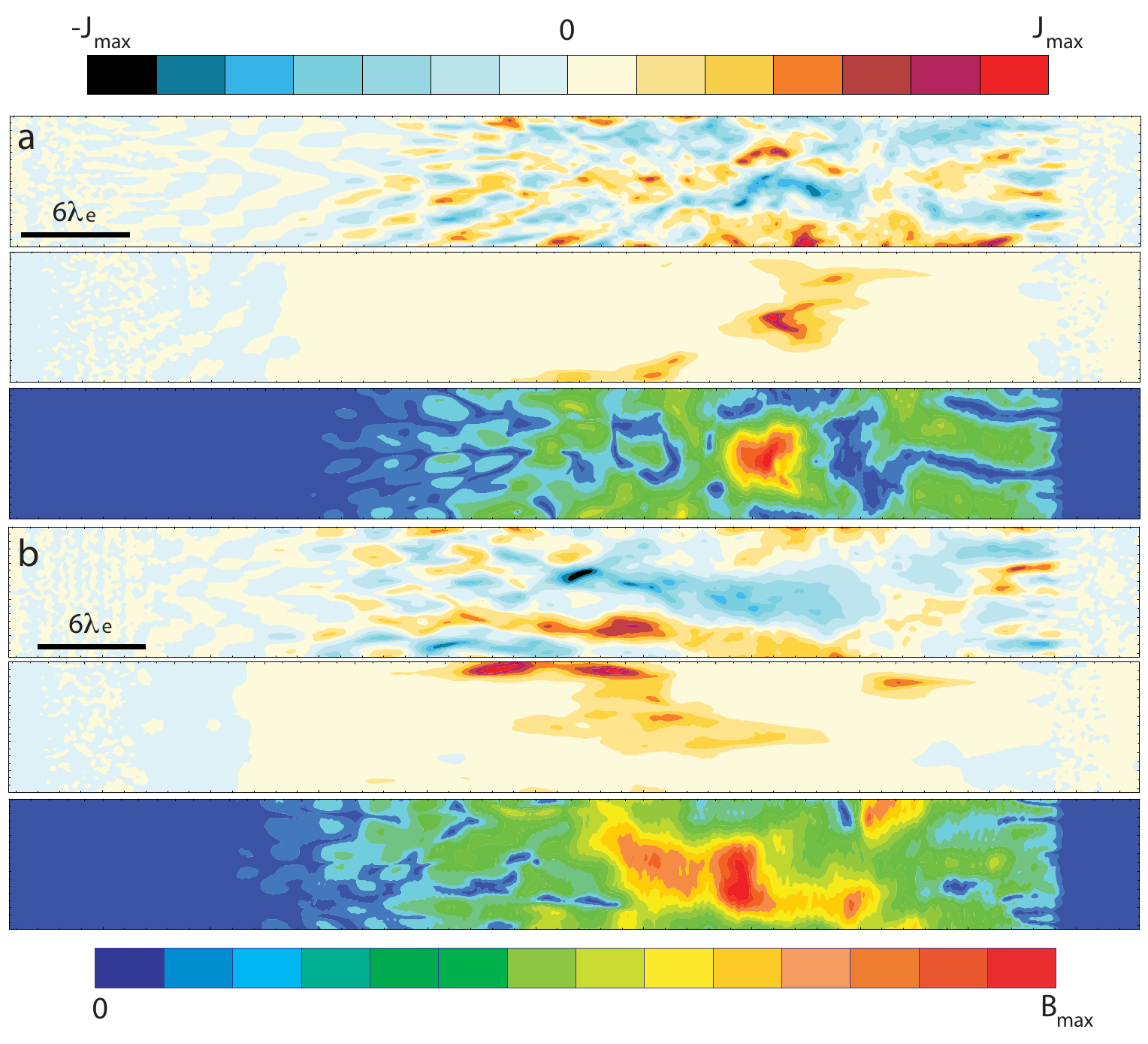}}
\caption{Current filamentation and magnetic field energy density after
  saturation: $t=59.8\omega_e^{-1}$. The amplitude of the average (in
  the {\it x-y} plane) transverse magnetic field, $B_\perp$ is plotted
  (in simulation units) as a function of $z$ together with the average
  $z$ component of the total current density, $J_z$ ({\it top}
  panel). The $z$ component of the ion current density is plotted for
  a subsection of the computational box: $y=43\Lambda$ ({\it middle}
  panel). The {\it a} ({\it b}) panels are for a simulation in which
  the initial momentum distribution function is monoenergetic
  (broadband). Color bar gives the amplitude of $J_z=[J_{\rm
      max},-J_{\rm max}]$ and $B_\perp=[0,B_{\rm max}]$ in simulation
  units: $J_{\rm max}=24.6$ (22.0), $J_{\rm max}=15.1$ (7.8) for
  ambient ions and $B_{\rm max}=3.6$ (2.8).}
\label{fig9}
\end{figure*}

From the requirement that the total plasma momentum should be
conserved, the electromagnetic field produced by the two-stream
instability acquires part of the longitudinal momentum lost by the
counterstreaming populations. Consequently, the magnetic field energy
grows in the early stage by slowing down the $e^{\pm}$ pair plasma, as
illustrated in Fig.~\ref{fig2}.  Our experiments show that this also
happens for $e^\pm$ pair shells injected into an electron-ion ambient
plasma (Fig~\ref{fig7}). Although as argued in \S \ref{pairs}, there
are some clear differences in the evolution of the associated
absorption of momentum between the broadband and the monoenergetic
cases. Figs.~\ref{fig8} and \ref{fig9} elucidate some of these
differences.

As can be clearly seen in Fig.~\ref{fig8}, transverse energy spreading
occurs faster in the broadband case, and, as a consequence, the
ambient ion beam is heated more promptly. The total magnetic energy
then grows as the ion channels merge.  The magnetic field associated
with these currents has also a filamentary structure, as seen in
Fig.~\ref{fig9}. The magnetic energy scales with the square of the
electric current, which in turn grows in inverse proportion to the
number of current channels. The net result is that the mean magnetic
energy increases accordingly: the magnetic field energy density
reaches 6.1\% (4.7\%) of the initial total kinetic energy in the
monoenergetic (broadband) case.

The ambient ions retain distinct bulk speeds in shielded ion channels
and thermalize much more slowly (Fig.~\ref{fig8}). This is expected to
continue as long as a surplus of bulk relative momentum remains in the
counterstreaming plasmas. Fig.~\ref{fig9} shows the extension of the
ion currents and the corresponding magnetic field energy
density. After the linear stage, the instability saturates and the
energy in the magnetic field decays, reflecting essentially the
subsequent decrease in the field's volume filling factor. In the
monoenergetic case, the magnetic energy density drastically drops
after about $\sim 30\omega_e^{-1}$. The decrease in the magnetic field
energy is, however, slower in the broadband shell. This is associated
with the continuous transverse energy spreading of progressively
faster moving $e^\pm$ pairs. The transverse energy spreading is less
efficient for the faster moving component and, as a result, a large
number of oppositely directed current filaments continue to be
generated even after the bulk of the flow enters the saturation
phase. In the monoenergetic case, on the other hand, the magnetic
energy density drastically drops after about $\sim
30\omega_e^{-1}$. These fields, however, maintain a strong saturated
level for at least the duration of the simulations. The final magnetic
energy density level is still quite high, up to 2.75\% (2.57\% ) of
the initial total kinetic energy in the monoenergetic (broadband)
case.

This is in contrast to the results reported by \cite{f04} for
relativistically counterstreaming electron-ion plasma collisions,
where the magnetic field energy density continues to grow throughout
their experiment, which lasts for about $480\omega_e^{-1}$.  In these
experiments, both counterstreaming plasmas are, however, composed of
electron and ions, which in turn significantly increases the relative
bulk momentum. It is thus clear that the symmetry associated with the
counterstreaming electron-ion populations is essential for the
longevity of the ion current channels.

\section{Conclusions}\label{con}

We present self-consistent three-dimensional simulations of the fields
developed by the electromagnetic filamentation instability as a
consequence of $e^\pm$ pair injection, which arise naturally in the
environment of GRB sources as a result of back-scattering. Our results
demonstrate that even in an initially unmagnetized scattering plasma,
a small-scale, fluctuating, predominantly transversal, and
near-equipartition magnetic field is unavoidably generated. These
fields maintain a strong saturated level on timescales much longer
than $\lambda_e$ at least for the duration of the simulations $\sim 60
(c/\lambda_e) \sim 0.1 (c/\lambda_T)$. The $e^\pm$ pairs are
effectively scattered with the magnetic field, thus effectively
communicated their momentum to the scattering medium initially at
rest. Our results indicate that the fields necessary to ensure that
$e^\pm$ pairs remain coupled to the medium can be easily created via
plasma instabilities. The next required step is to increase the
ion-to-electron mass ratio in the scattering medium in order to
determine the spatial spread and character of the particle coupling.

A question which has remained largely unanswered so far, is what
determines the characteristic strength of the upstream magnetic field
in afterglow shocks, which is inferred to extend to tenths, or even a
tens, of mG \citep{lw06}. This is of course large in comparison with
the magnetic fields presumed to be present in the interstellar medium,
which are measured in $\mu$ G. A sufficiently strong magnetic field
may be present in a wind ejected by the GRB progenitor but only if its
is highly magnetized, which is not thought to be the case for the
widely favored Wolf-Rayet stars \citep{lw06}. The injection of $e^\pm$
pairs induces strong streaming motions in the ambient medium ahead of
the forward shock. This sheared flow, as we demonstrated here, will be
Weibel-like unstable and, if $e^{\pm}$ loading continues, well after
the shock hits the $e^{\pm}$ pair enriched medium, the resulting
upstream flow will be strongly magnetized. Since one expects the
radiation front to lead the forward shock \citep{b02,b05} by a small
distance, $R/(4\Gamma^2)$, one expects the instability to be able to
generate subequipartition magnetic fields, $B\sim 0.1(n_p/1\;{\rm
  cm^{-3}})^{1/2}$ G, in the scattering medium just before it is
shocked. In this case, the constraint on the upstream field strength
of $B\gg 0.05 (n_p/1\;{\rm cm^{-3}})^{5/8}$~mG, imposed by X-ray
afterglow observations \citep{lw06} is easily satisfied.

\acknowledgments We have benefited from many useful discussions with
J. Arons, P. Chang, M. Rees, C. Thompson and E. Waxman. We are
especially grateful to U. Keshet for countless insightful
conversations. We also thank the referee for many useful comments that
helped improve the paper.  This work is supported by NSF: PHY-0503584
(ER-R), ATM 9730230, ATM-9870072, ATM-0100997, INT-9981508,
AST-0506719, NASA-NNG05GK73G (KN) and DOE SciDAC: DE-FC02-01ER41176
(ER-R). Simulations were performed on Columbia at NASA Advanced
Supercomputing (NAS) and IBM p690 (Copper) at the National Center for
Supercomputing Applications (NCSA) which is supported by the NSF. Part
of this work was done while ER-R was visiting the DARK and MSFC. ER-R
thank the directors of these institutions for their generous
hospitality.

\newpage

\begin{thebibliography}{}

\bibitem[Beloborodov(2002)]{b02} Beloborodov, A.~M.\ 2002, \apj, 565,
808

\bibitem[Beloborodov(2005)]{b05} Beloborodov, A.~M.\ 2005, ApJ, 627,
346

\bibitem[Buneman(1993)]{b93} Buneman, O., 1993, Tristan, in Computer
Space Plasma Physics: Simulation Techniques and Software, edited by
H.\ Matsumoto Matsumoto \& Y.\ Omura, p. 67, Terra Scientific
Publishing Company, Tokyo

\bibitem[Frederiksen et al.(2004)]{f04} Frederiksen, J.~T. et
al. 2004, ApJ, 608, L13

\bibitem[Gruzinov(2001)]{g01} Gruzinov, A. 2001, ApJ, 563, L15

\bibitem[Hededal \& Nishikawa(2005)]{hn05} Hededal, C.~B., \&
Nishikawa, K.-I.\ 2005, ApJ, 623, L89

\bibitem[Jaroschek et al.(2005)]{jaro05} Jaroschek, C. H., Lesch, H.,
\& Treumann, R. A. 2005, ApJ, 618, 822

\bibitem[Jelley(1966)]{jel66} Jelley, J. V. 1966, Nature, 211, 472

\bibitem[Kumar \& Panaitescu(2004)]{kp04} Kumar, P., \& Panaitescu,
A.\ 2004, MNRAS, 354, 252

\bibitem[Li et al.(2003)]{li03} Li, Z. et al. 2003, ApJ, 599, 380

\bibitem[Li \& Waxman(2006)]{lw06} Li, Z., \& Waxman, E.\ 2006, ApJ in
press, astro-ph/0603427

\bibitem[Medvedev \& Loeb(1999)]{ml99} Medvedev, M. V., \& Loeb,
A. 1999, ApJ, 526, 697

\bibitem[M{\'e}sz{\'a}ros et al.(2001)]{m01} M{\'e}sz{\'a}ros, P.,
Ramirez-Ruiz, E., \& Rees, M.~J.\ 2001, \apj, 554, 660

\bibitem[Nishikawa et al.(2005)]{n05} Nishikawa, K.-I. et al. 2005,
ApJ, 622, 927

\bibitem[Nishikawa et al.(2006)]{n06} Nishikawa, K.-I. et al. 2006,
ApJ, 642, 1267

\bibitem[Piran(1999)]{p99} Piran, T. 1999, Phys. Rep., 314, 575

\bibitem[Pruet et al.(2001)]{p01} Pruet, J., Abazajian, K., Fuller,
G. M. 2001, Phys.\ Rev.\ D., 64, 063002

\bibitem[Ramirez-Ruiz et al.(2002)]{rr02} Ramirez-Ruiz, E., MacFadyen,
A.~I., \& Lazzati, D.\ 2002, MNRAS, 331, 197

\bibitem[Silva et al.(2002)]{s02} Silva, L. O., et al. 2002,
Phys. Plasmas, 9, 2458

\bibitem[Silva et al.(2003)]{s03} Silva, L.~O., Fonseca, R.~A.,
Tonge, J.~W., Dawson, J.~M., Mori, W.~B., \& Medvedev, M.~V.\ 2003,
ApJ, 596, L121

\bibitem[Spitkovsy (2006)]{s06} Spitkovsky, A. 2006, astro-ph/0603211

\bibitem[Svensson(1987)]{sve87} Svensson, R. 1987, MNRAS, 227, 403

\bibitem[Thompson \& Madau (2000)]{tm00} Thompson, C., \& Madau,
P. 2000, ApJ, 538, 105



\end{thebibliography}
\end{document}